\documentclass[english]{article}
\usepackage[T1]{fontenc}
\usepackage[latin9]{luainputenc}
\usepackage{geometry}
\usepackage{amstext}
\usepackage{amssymb}
\usepackage{feyn}
\usepackage{tikz}
\usepackage{graphicx}
\usepackage{array}
\usepackage{booktabs}

\usepackage{amsmath}
\usepackage{amsfonts}
\usepackage{mathrsfs}

\usepackage[normalem]{ulem}
\usepackage{color}
\usepackage{listings,braket}
\usepackage{caption}
\usepackage{subcaption}
\usepackage{float}
\definecolor{darkgreen}{rgb}{0,0.35,0}
\definecolor{Rood}{rgb}{1, 0, 0}

\makeatletter
\usepackage[english]{babel}
\usepackage{feyn}
\usepackage{simpler-wick}

\makeatother

\begin{document}

	\title{\bf Useful Trick to Compute Correlation Functions of Composite Operators}

	{\author{ \textbf{Giovani~Peruzzo$^1$}\thanks{gperuzzofisica@gmail.com}
			\\\\\
			\textit{{\small $^1$Instituto de F\'{i}sica, Universidade Federal Fluminense,
			}}\\
			\textit{{\small Campus da Praia Vermelha, Av. Litor\^{a}nea s/n, }}\\
			\textit{{\small 24210-346, Niter\'{o}i, RJ, Brasil}}\\
			\\
		}
		
		\date{}
		
		\maketitle

\begin{abstract}
In general, in gauge field theories, physical observables are represented
by gauge-invariant composite operators, such as the electromagnetic
current. As we recently demonstrated in the context of the $U\left(1\right)$
and $SU\left(2\right)$ Higgs models \cite{Dudal:2019pyg,Dudal:2020uwb,Maas:2020kda}, correlation functions
of gauge-invariant operators exhibit very nice properties. Besides the
well-known gauge independence, they do not present unphysical cuts,
and their K\"{a}ll\'{e}n-Lehmann representations are positive, at least perturbatively.
Despite all these interesting features, they are not employed as much
as elementary fields, mainly due to the additional complexities
involved in their computation and renormalization. In this article,
we present a useful trick to compute loop corrections to correlation
functions of composite operators. This trick consists of introducing an additional
field with no dynamics, coupled to the composite operator of interest.
By using this approach, we can employ the traditional
algorithms used to compute correlation functions of elementary fields.
\end{abstract}

\section{Introduction}
In Quantum Field Theory (QFT), in addition to the correlation functions of elementary fields, we are also interested in the correlation functions of composite operators, which are formed by the product of fields at the same point. Currents are a prime example of composite operators. These objects play a crucial role in theories where the symmetry content is highly significant, such as gauge field theories. Moreover, these operators are fundamental to the Operator-Product Expansion (OPE) technique \cite{Wilson:1969zs,Collins:1984xc}, which is applied in various contexts, including QCD (Quantum Chromodynamics) sum rules, deep inelastic scattering, the Wilsonian Renormalization Group, and more. Composite operators are also fundamental to the implementation of non-linear symmetry at the quantum level, such as the BRST symmetry \cite{Becchi:1974xu,Tyutin:1975qk}. \par 
Recently, interest in composite field theory has been reignited in the context of gauge theories, specifically in composite gauge-invariant operators, due to their spectral and gauge-independent properties \cite{Dudal:2019pyg,Dudal:2020uwb,Maas:2020kda,Kugo:1979gm}. At least formally, the correlation functions of gauge-invariant operators are gauge independent \cite{Nielsen:1975fs,Piguet:1984js}. In practice, however, this is more complex because, in non-Abelian theories, the well-known Gribov-Singer ambiguities arise \cite{Gribov:1977wm,Singer:1978dk}. Nonetheless, if we disregard these ambiguities and consider the perturbative expansion obtained \emph{via} the Faddeev-Popov method \cite{Faddeev:1967fc} or BRST quantization \cite{Becchi:1974xu}, the correlation functions of gauge-invariant operators are, at the very least, independent of any gauge parameter. Furthermore, the positivity of the K\"{a}ll\'{e}n-Lehmann representation for these operators stems from the decoupling of non-physical degrees of freedom, such as ghost fields, and the semi-positivity of the physical subspace defined by the cohomology of the BRST charge \cite{Kugo:1979gm}. These are two key examples of properties shared by these operators that enable a clear and direct understanding of the physical content of the underlying gauge theory.        
\par

Despite these interesting features, composite operators require special treatment when it comes to renormalization, which may discourage their use. In general, for a correlation function such as $\langle \phi\left(x\right)\phi\left(y\right) \ldots \rangle$, the limit $x \to y$ is not well-defined, even in renormalizable theories. Consequently, a precise and distinct definition for the composite operator $\phi^2\left(x\right)$ must be established from the outset. The key issue is that composite operators introduce new types of UV divergences that cannot be eliminated by the counterterms used for renormalizing the correlation functions of elementary fields. However, there is a well-established framework to address this problem (see \cite{Collins:1984xc,Itzykson:1980rh}). While renormalization may be cumbersome, an algorithm exists to systematically handle it. Another reason some may avoid composite operators is computational practicality. It is generally easier to implement computations involving elementary fields in software, such as Mathematica, than those involving composite operators. The primary goal of this article is to introduce an alternative method - essentially a computational trick - for obtaining correlation functions of composite operators in terms of the correlation functions of an auxiliary elementary field, which we conveniently add to the theory. With minimal adaptations, this approach allows us to apply the same algorithms used for elementary fields to compute the correlation functions of composite operators. \par

The basic idea involves introducing an auxiliary field, denoted by $B(x)$, which is coupled to the composite operator of interest, say $O(x)$, along with a new parameter $\alpha$. This auxiliary field is non-dynamical and, on-shell, corresponds to $O(x)$. We demonstrate that the diagrammatic structure of the modified theory, which includes the interaction vertex $B(x)O(x)$, is consistent with the diagrammatic structure of the composite operator $O(x)$. More specifically, we show that the $\alpha^r$-order of $\langle B(x_1) \ldots B(x_r) \rangle_c$ in this modified theory corresponds to the correlation function $\langle O(x_1) \ldots O(x_r) \rangle_c$ in the original theory. This correspondence constitutes the main result of this article. We test this method for the $\phi^2(x)$ operator in the $\lambda\phi^4$ theory and show that it works, at least formally. In the final section, we provide strong arguments suggesting that this approach may hold at all orders in perturbation theory, even when the theory used to map the original problem is not renormalizable. \par 

The article is organized as follows. In Section \ref{Brief review of composite operators}, we present a brief review of composite operators. In Section \ref{sec:Introducing-the-new}, the method is presented. In Section \ref{Demonstrating the Trick in Practice}, we apply this method to the operator $\phi^2(x)$ in the $\lambda \phi^4$ theory. In Section \ref{Final Remarks}, we discuss the perturbative consistency and the renormalization of the theory with the field $B\left(x\right)$.

\section{A Brief Overview of Composite Operators}\label{Brief review of composite operators}

Let $O\left(x\right)$ be a composite field obtained by the product
of $n$ elementary real scalar fields $\phi$ at the same point $x$, \emph{i.e.},
\begin{eqnarray}
O\left(x\right) & = & \frac{1}{n!}\phi^{n}\left(x\right). \label{eq:On}
\end{eqnarray}
This means that, from a certain external point, say $x$, we have
$n$ fields of $\phi$ to be contracted with other fields. According
to the Wick theorem, all fields must be contracted two by two to form
lines. Therefore, from a point with $O$, there are no more than $n$
lines attached, as we have to consider that might be contractions
among the fields of $O$, as shown in Figure 1. Hence, for many practical,
a composite operator can be consider as a new vertex, but a non-integrated
one, since it depends on an external point. Thus, if we go to the
momentum space, we do not obtain the momentum conservation in the
vertice containing the composite operator, \emph{i.e.} 
\begin{gather}
p_{1}+\ldots+p_{n}\equiv-p\neq0.\label{eq:imaginary_external_momenta}
\end{gather}
However, if we consider $p$ as the external momenta flowing in the
vertex, see Figure 2, we reobtain the usual prescription for computing
Feynman diagrams. These observations lead us, fundamentally, to Feynman's
rules for calculating correlation functions of composite operators.
\begin{figure}
\begin{centering}
\includegraphics[scale=0.25]{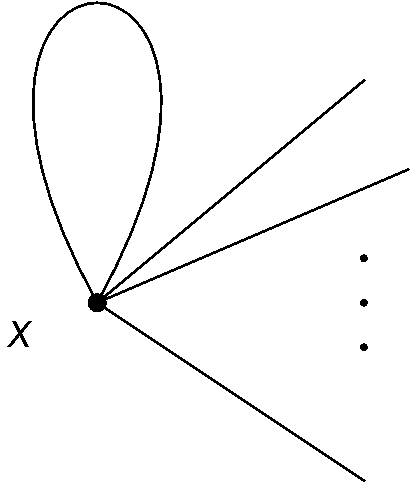}
\par\end{centering}
\caption{Diagram with one contraction between fields of the same $O\left(x\right)$.}

\end{figure}

\begin{figure}
\begin{centering}
\includegraphics[scale=0.25]{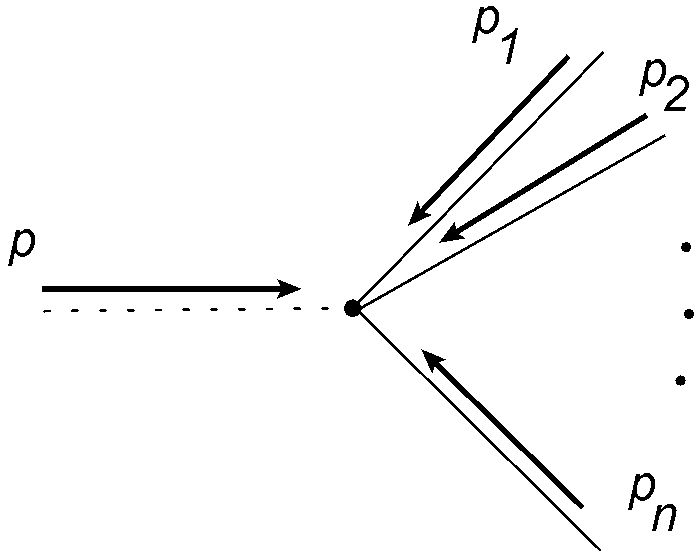}
\par\end{centering}
\caption{From the diagrammatic perspective, a local composite operator can
be consider as a vertex with an additional external momenta $p$,
which can be associated to an imaginary external line. }

\end{figure}

Due to the need of renormalization, we have introduce $O$ coupled
with an external source, say $\rho$, and modify the starting action
$S\left[\phi\right]$ to
\begin{gather}
S\left[\phi\right]\rightarrow S\left[\phi\right]+\int d^{d}x\rho\left(x\right)O\left(x\right). \label{O_external_source}
\end{gather}
Consequently, besides the usual source $J\left(x\right)$ used to
introduce $\phi\left(x\right)$, the generating functional of correlation
functions will be a functional of $\rho$ too, namely,
\begin{eqnarray}
Z\left[J,\rho\right] & = & \frac{\int D\phi\,e^{-\left(S+\int d^{d}xJ\phi+\int d^{d}x\rho O\right)}}{\int D\phi\,e^{-S}},\label{eq:Z_functional}
\end{eqnarray}
Therefore, the correlation functions of $O$ are defined as
\begin{eqnarray}
\left\langle O\left(x_{1}\right)\ldots O\left(x_{r}\right)\right\rangle ^{S} & = & \left(-1\right)^{r}\left.\frac{\delta^{r}Z\left[J,\rho\right]}{\delta\rho\left(x_{1}\right)\ldots\delta\rho\left(x_{r}\right)}\right|_{J=\rho=0}.\label{eq:O_O}
\end{eqnarray}
This formulation allows us to treat the renormalization of $\left\langle O\left(x_{1}\right)\ldots O\left(x_{r}\right)\right\rangle $
systematically, as any counterterm necessary to renormalize it can
be obtained in the same way, \emph{i.e.}, as functional derivatives
of $\rho$. 

\section{A Novel Method for Correlation Functions of Composite Operators}\label{sec:Introducing-the-new}

In the previous section, we described the basic Feynman rules to
compute Feynman diagram of composite operators. As we discussed, $O=\frac{1}{n!}\phi^{n}$
behaves like a vertex with $n$ lines and an addtional momenta $p$.
This kind of vertex is found in a theory with the coupling $\frac{1}{n!}B\phi^{n}$,
where $B$ is a new real scalar field. Since there is no external
legs attached to the composite operator vertex, only the external
momenta $p$, it is convenient to take $B$ as a non-propagating field,
which means that its propagator in momentum space is constant. All
these features are obtained in the theory defined by the action
\begin{gather}
S_{B}=S+\int d^{d}x\left(\frac{B^{2}}{2}-\alpha\frac{1}{n!}B\phi^{n}\right). \label{eq:Sb}
\end{gather}
Indeed, we do have the constant propagator of $B$, namely,
\begin{gather}
\Delta_{BB}\left(p\right) =  1\Leftrightarrow\Delta_{BB}\left(x-y\right)=\delta\left(x-y\right),\label{eq:BB_prop}
\end{gather}
and the $B\phi$-vertex we want. Notice that we have introduced a
new parameter $\alpha$, the meaning of which will become clear shortly. As we mentioned in the Introduction, $B\left(x\right)$ on-shell corresponds to the operator $O\left(x\right)$, \emph{i.e.},
\begin{gather}
	\frac{\delta S_{B}}{\delta B}=B-\alpha \frac{1}{n!}\phi^n=0 \Leftrightarrow\ B=-\alpha \frac{1}{n!}\phi^n.
\end{gather}

Now, let us analyze the theory described by $S_{B}$. As we argued,
there is a relationship, that we aim to clarify, between correlation
functions $O$, see Eq. (\ref{eq:O_O}), and the correlation functions
of $B$, defined as
\begin{gather}
\left\langle B\left(x_{1}\right)\ldots B\left(x_{r}\right)\right\rangle ^{S_{B}}=\frac{\int D\phi DB\,B\left(x_{1}\right)\ldots B\left(x_{r}\right)e^{-S_{B}}}{\int DB\,e^{-S_{B}}}.
\end{gather}
Given that only one vertex with $B$ exists, namely, $\int d^{d}x\alpha\frac{1}{n!}B\phi^{n}$,
there is no other option for a certain external $B$ besides to contract
with this vertex or another external $B$. In general, the second
option produces a disconnected diagram, which does not arise in the
case of a composite operator. Therefore, to exclude this possibility,
we will henceforth consider only connected correlation functions,
indicated by the label ``c'', and connected diagrams. Certainly,
this does not represents a limitation in our analysis, as we can always
use connected correlation function to construct correlation function.
Thus, the only remaining case is the first one, when all external
$B$'s contract with the interaction vertex involving $B$ and $\phi$.
In this case, as each contraction of $B$'s produces a Dirac delta
function according to (\ref{eq:BB_prop}), it follows that 
\begin{gather*}
\wick{\c B\left(x\right)\ldots\left(\int d^{d}y\alpha\frac{1}{n!}\c B\left(y\right)\phi^{n}\left(y\right)\right)}\ldots=\int d^{d}y\delta^{d}\left(x-y\right)\alpha\frac{1}{n!}\phi^{n}\left(y\right)\ldots=\alpha\frac{1}{n!}\phi^{n}\left(x\right)\ldots,
\end{gather*}
meaning that, effectively, we have the composite operator $\frac{1}{n!}\phi^{n}\left(x\right)$
in each external point.

The first correction to a $r$-point function of $B$ will be of order
$\alpha^{r}$, corresponding to the case where all $r\cdot n$ internal
lines originating from the $r$ vertices contract among themselves.
Here is an important observation: this is exactly the same correction
we obtain for $\left\langle O\left(x_{1}\right)\ldots O\left(x_{r}\right)\right\rangle _{c}^{S}$
when we set $\alpha=1$. Higher-order corrections in $\alpha$ to
$\left\langle B\left(x_{1}\right)\ldots B\left(x_{r}\right)\right\rangle _{c}^{S_{B}}$
necessarily involve more than $r$ $B\phi$-vertices, so they must
be discarded, as the number of composite operator is always fixed
at $r$. Therefore, we can conclude that $\left\langle O\left(x_{1}\right)\ldots O\left(x_{r}\right)\right\rangle _{c}^{S}$
is encoded in the order-$\alpha^{r}$ correction (which is also the
first perturbative order) of $\left\langle B\left(x_{1}\right)\ldots B\left(x_{r}\right)\right\rangle _{c}^{S_{B}}$.
Since perturbative correction involving other couplings of $S$ are
the same, we establish that
\begin{eqnarray}
\left\langle O\left(x_{1}\right)\ldots O\left(x_{r}\right)\right\rangle _{c}^{S} & = & \frac{1}{r!} \left.\frac{\partial^{r}}{\partial\alpha^{r}}\left\langle B\left(x_{1}\right)\ldots B\left(x_{r}\right)\right\rangle _{c}^{S_{B}}\right|_{\alpha=0}. \label{eq:master_equation}
\end{eqnarray}
\par 
In Appendix A, we present a different demonstration of the result \eqref{eq:master_equation} using a functional method that relies on a Ward identity. Therefore, we provide a solid mathematical argument on which \eqref{eq:master_equation} can be based. This Ward identity keeps the UV divergences under control by establishing constraints on the counterterms necessary to renormalize $\langle O\left(x_1\right)\dots O\left(x_r\right)\rangle^S_c$ and $\langle B\left(x_1\right)\dots B\left(x_r\right)\rangle^{S_B}_c$, which implies that \eqref{eq:master_equation} holds not only for the regularized correlation functions but also for the renormalized ones.

\section{Demonstrating the Trick in Practice}\label{Demonstrating the Trick in Practice}

Having outlined the general ideas, let us now observe the trick in
action. Here, we consider the example of $\lambda\phi^{4}$ theory,
characterized by the action
\begin{eqnarray}
S & = & \int d^{d}x\left(\frac{1}{2}\partial_{\mu}\phi\partial_{\mu}\phi+\frac{m^{2}}{2}\phi^{2}+\frac{\lambda}{4!}\phi^{4}\right)\label{eq:phi_4_action}
\end{eqnarray}
and the quadratic composite operator
\begin{eqnarray}
O\grave{\left(x\right)} & = & \frac{\phi^{2}\left(x\right)}{2}.
\end{eqnarray}
$m$ and $\lambda$ are the massive and quartic couplings of the theory,
respectively. This is a well-known laboratory for testing and presenting
ideas in Quantum Field Theory, thus it is unnecessary to say many
things about this model. Regarding the quadratic operator, we can
mention that, at the classical level, this object is related to the
Conformal Symmetry breaking, see \cite{Collins:1984xc}. From the classical
action (\ref{eq:phi_4_action}) we derive the following propagator
and vertex\footnote{We use the following  definition for propagators and vertices: $\left\langle \Phi\left(x\right)\Phi\left(0\right)\right\rangle =\int\frac{d^{4}p}{\left(2\pi\right)^{d}}e^{-ip\cdot x}\Delta_{\Phi\Phi}\left(p\right)$
and $\left(2\pi\right)^{d}\delta^{d}\left(p_{1}+\ldots+p_{n}\right)\Gamma_{\Phi\ldots\Phi}\left(p_{1},\ldots,p_{n}\right)=-\prod_{i=1}^{n}\int d^{d}x_{i}e^{ip_{i}\cdot x_{i}}\left.\frac{\delta^{n}S}{\delta\Phi\left(x_{1}\right)\ldots\delta\Phi\left(x_{n}\right)}\right|_{\Phi=0}$.}, respectively:
\begin{eqnarray}
\Delta_{\phi\phi}\left(p\right) & = & \frac{1}{p^{2}+m^{2}},\nonumber \\
\Gamma_{\phi\phi\phi\phi}\left(p_{1},p_{2},p_{3},p_{4}\right) & = & -\lambda,\label{eq:feynman_rules_S}
\end{eqnarray}
The diagrammatic representation of these object are shown in Figure
3.
\begin{figure}
\begin{centering}
\includegraphics[scale=0.25]{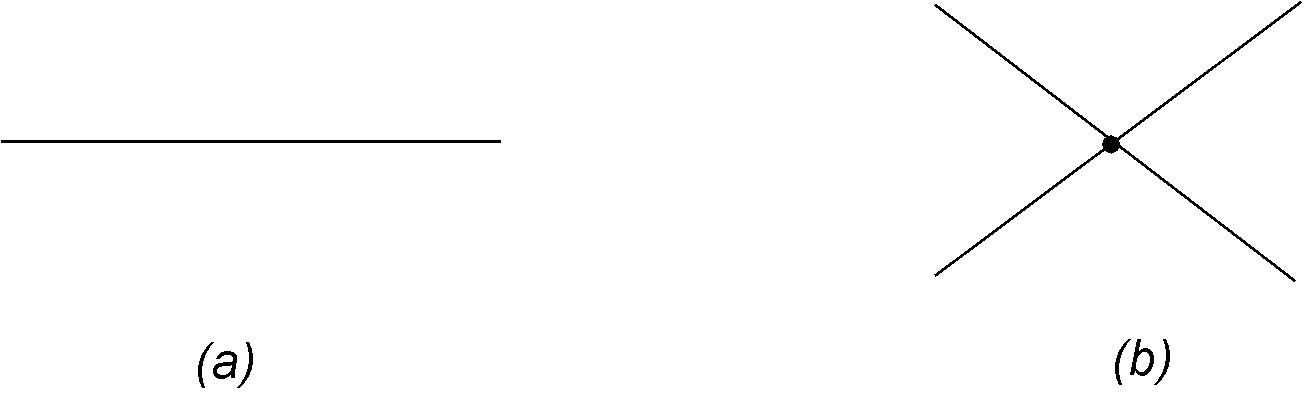}
\par\end{centering}
\caption{Diagrammatic representation of (a) the propagator of $\phi$ and (b)
the quadratic vertex. }

\end{figure}

Some of the Feynman diagrams that contribute to the connected two-point
function of $O$ are shown in Figure 4. Notice that their diagrams
start from one-loop onward. This is well-known property of pure composite
operators. If $O$ would be a linear combination of a composite and
a linear operator, there would also be a tree-level contribution,
see \cite{Dudal:2019pyg,Capri:2020ppe,Dudal:2021dec,Dudal:2023jsu,Peruzzo:2024heb}. By reading of these diagrams using the Feynman
rules, we obtain the following mathematical expression for $\left\langle O\left(x\right)O\left(y\right)\right\rangle _{c}^{S}$
in momentum space:
\begin{gather}
\left\langle O\left(p\right)O\left(-p\right)\right\rangle _{c}^{S} =  \frac{1}{2}\int\frac{d^{d}k}{\left(2\pi\right)^{d}}\Delta_{\phi\phi}\left(k\right)\Delta_{\phi\phi}\left(p-k\right)-\frac{\lambda}{4}\int\frac{d^{d}k}{\left(2\pi\right)^{d}}\Delta_{\phi\phi}\left(k\right)\Delta_{\phi\phi}\left(p-k\right)\int\frac{d^{d}l}{\left(2\pi\right)^{d}}\Delta_{\phi\phi}\left(l\right)\Delta_{\phi\phi}\left(p-l\right) \nonumber\\
-\frac{\lambda}{4}\int\frac{d^{d}k}{\left(2\pi\right)^{d}}\Delta_{\phi\phi}\left(k\right)^{2}\Delta_{\phi\phi}\left(k-p\right)\int\frac{d^{d}l}{\left(2\pi\right)^{d}}\Delta_{\phi\phi}\left(l\right)\nonumber\\
  +\frac{\lambda^2}{8}\int\frac{d^{d}k}{\left(2\pi\right)^{d}}\Delta_{\phi\phi}\left(k\right)\Delta_{\phi\phi}\left(p-k\right)\int\frac{d^{d}l}{\left(2\pi\right)^{d}}\Delta_{\phi\phi}\left(l\right)\Delta_{\phi\phi}\left(k-l\right)\int\frac{d^{d}q}{\left(2\pi\right)^{d}}\Delta_{\phi\phi}\left(q\right)\Delta_{\phi\phi}\left(p-q\right)+\ldots \label{eq:OO_expression}
\end{gather}

\begin{figure}
\begin{centering}
\includegraphics[scale=0.16]{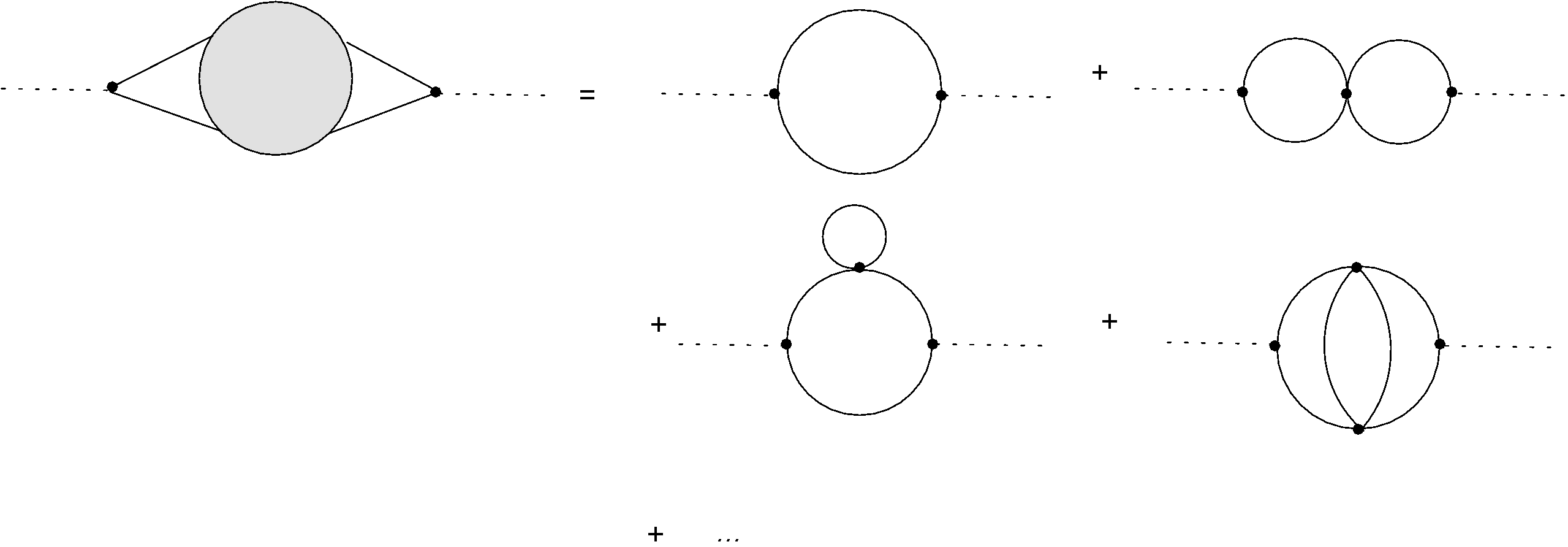}
\par\end{centering}
\caption{Feynman diagrams that contribute to $\left\langle O\left(x\right)O\left(y\right)\right\rangle _{c}^{S}$.}

\end{figure}

Now, let us apply the trick described in Section \ref{sec:Introducing-the-new},
to calculate the same correlation function. By following the prescription,
we consider the correlation function $\left\langle B\left(x\right)B\left(y\right)\right\rangle _{c}^{S_{B}}$
in the theory defined by the action
\begin{eqnarray*}
S_{B} & = & S+\int d^{d}x\left(\frac{B^{2}}{2}-\alpha\frac{B\phi^{2}}{2}\right)\\
 & = & \int d^{d}x\left(\frac{1}{2}\partial_{\mu}\phi\partial_{\mu}\phi+\frac{m^{2}}{2}\phi^{2}+\frac{\lambda}{4!}\phi^{4}+\frac{B^{2}}{2}-\alpha\frac{B\phi^{2}}{2}\right).
\end{eqnarray*}
Additionaly to the Feynman rules (\ref{eq:feynman_rules_S}), we have
the propagator of $B$, see Eq. (\ref{eq:BB_prop}), and the vertex
\begin{eqnarray*}
\Gamma_{B\phi\phi}\left(p_{1},p_{2},p_{3}\right) & = & \alpha,
\end{eqnarray*}
whose diagrammatic representations are shown in Figure 5. 
\begin{figure}
\begin{centering}
\includegraphics[scale=0.25]{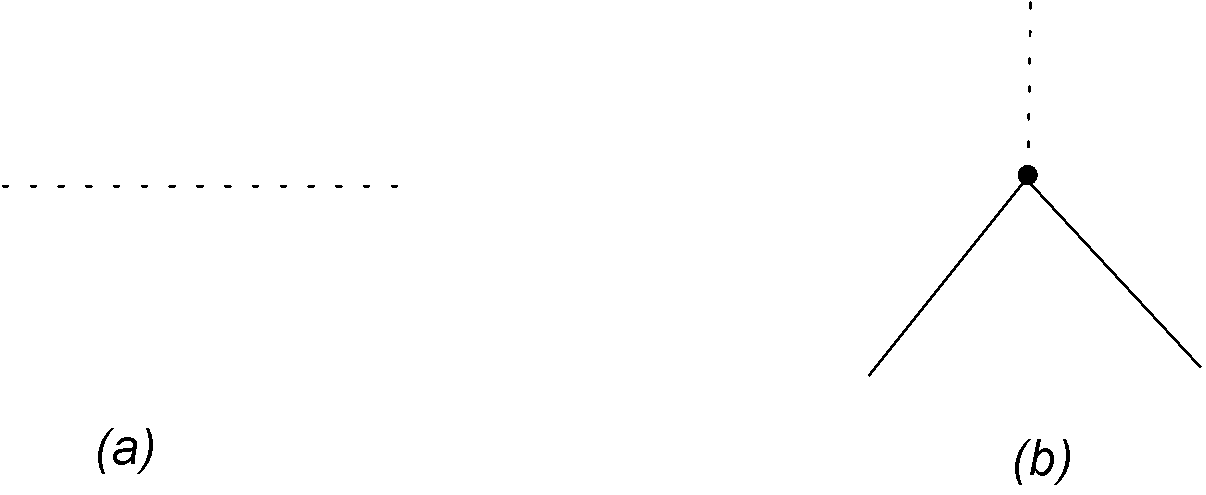}
\par\end{centering}
\caption{Diagrammatic representation of (a) the propagator of $B$ and (b)
the $B\phi$-vertex. }

\end{figure}
Notice that, on purpose, we use the same line of the imaginary line
carrying the external momenta $p$ (see Eq. (\ref{eq:imaginary_external_momenta}))
to represent the propagator of $B$. 

In Figure 6, we present some of the Feynman diagrams that contribute
to the function $\left\langle B\left(x\right)B\left(y\right)\right\rangle _{c}^{S_{B}}$.
Alongside the diagrams, we indicate the order in $\alpha$. As
the reader can compare, the diagrams of $\alpha^{2}$-order are the same as those
of $\left\langle O\left(x\right)O\left(y\right)\right\rangle _{c}^{S}$,
see Figure 4. If we write down the corresponding mathematical expression, we also obtain the same result as \eqref{eq:OO_expression}, since the external lines are equal to 1. 

The next order of corrections in $\alpha$ involves diagrams with four $B\phi$-vertices. Consequently, each diagram includes one internal $B$ line. By applying the Feynman rules to the first diagram at $\alpha^4$-order shown in Figure 6, we obtain the result: 
\begin{gather}
	\frac{\alpha^4}{4}\Delta_{BB}(p)\int \frac{d^dk}{\left(2\pi\right)^d}\Delta_{\phi\phi}\left(k\right)\Delta_{\phi\phi}\left(p-k\right)\Delta_{BB}\left(p\right)\int \frac{d^dl}{\left(2\pi\right)^d}\Delta_{\phi\phi}\left(l\right)\Delta_{\phi\phi}\left(p-l\right)\Delta_{BB}\left(p\right) \nonumber \\
	= \frac{\alpha^4}{4}\int \frac{d^dk}{\left(2\pi\right)^d}\Delta_{\phi\phi}\left(k\right)\Delta_{\phi\phi}\left(p-k\right)\int \frac{d^dl}{\left(2\pi\right)^d}\Delta_{\phi\phi}\left(l\right)\Delta_{\phi\phi}\left(p-l\right)
\end{gather} 
Notice that we would obtain a similar expression if we had one $\phi^4$-type vertex instead of two $B\phi$-vertices, see the second term on the RHS of Eq. \eqref{eq:OO_expression}. This is a general property that can be directly inferred from the contraction of two $B$'s of two different vertices: 
\begin{gather}
	\wick{\int d^d x \frac{\alpha}{2}\c B(x)\phi^2(x) \int d^d y \frac{\alpha}{2}\c B(y)\phi^2(y) }=\int d^d x \frac{\alpha}{2}\phi^2(x) \int d^d y \frac{\alpha}{2}\delta^d\left(x-y\right)\phi^2(y)=\int d^d x \frac{\alpha^2}{4} \phi^4(x) \label{eq:Bphi2_phi4}
\end{gather}
The diagrammatic representation of the collapse \eqref{eq:Bphi2_phi4} is shown in Figure 7.

\begin{figure}
	\begin{centering}
		\includegraphics[scale=0.13]{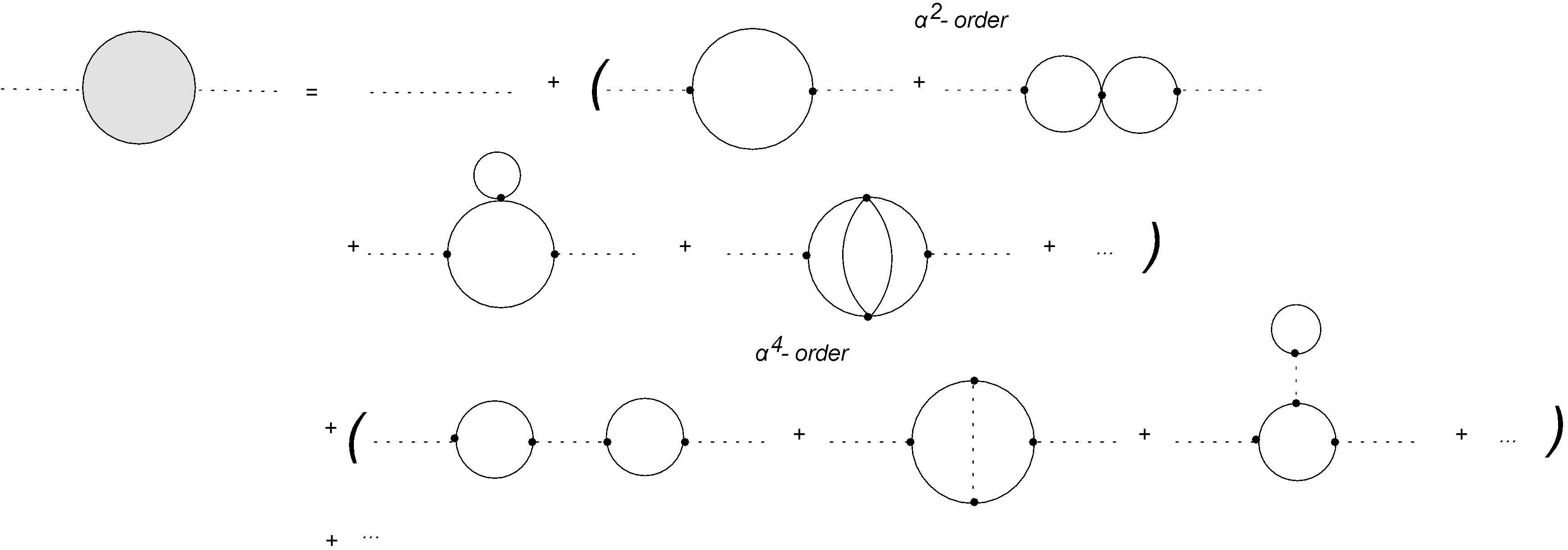}
		\par\end{centering}
	\caption{Diagrams that contribute to $\left\langle B\left(x\right)B\left(y\right)\right\rangle _{c}^{S_{B}}$
		with the respective order in $\alpha$.}
	
\end{figure}

\begin{figure}
	\begin{centering}
		\includegraphics[scale=0.15]{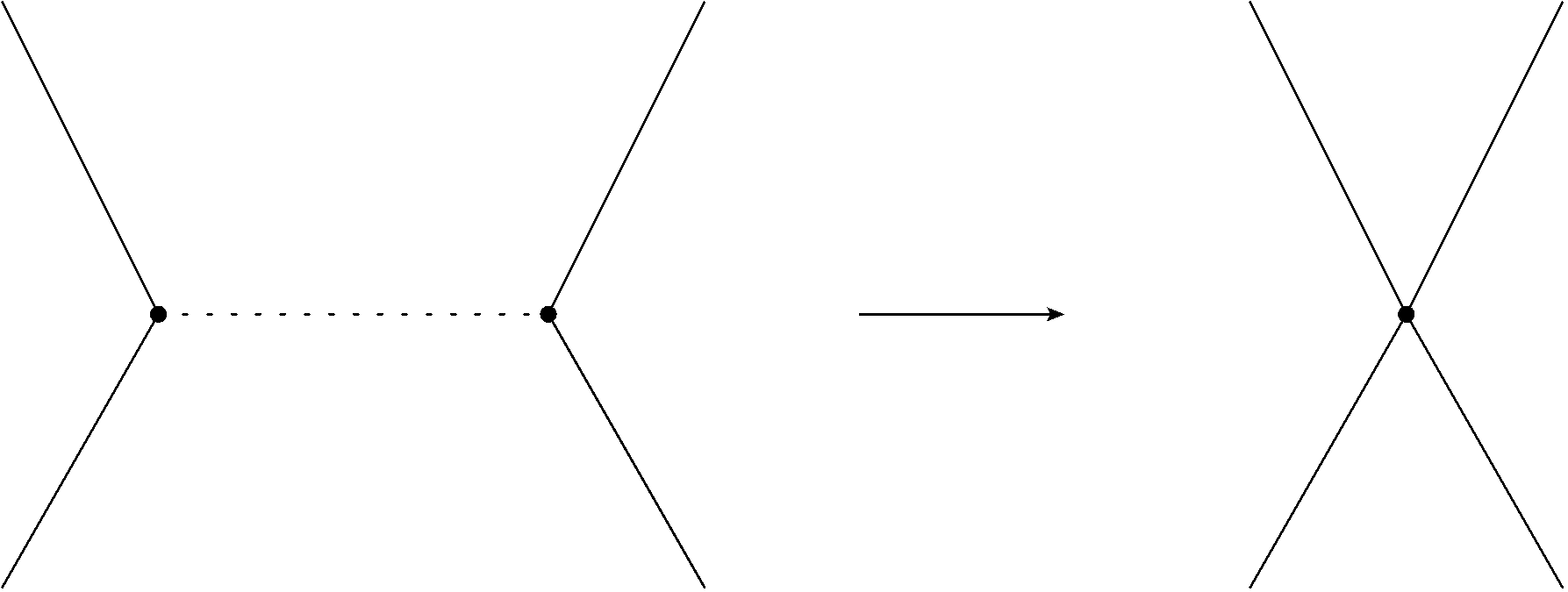}
		\par\end{centering}
	\caption{When two $B$'s of internal vertices contract, the result is equivalent to a $\phi^4$-type vertex.}
	
\end{figure}

\section{Final Remarks on the All-Orders Consistency of the Method}\label{Final Remarks}
Our main goal in this article is to present a useful method for computing correlation functions of composite operators, particularly within already implemented algorithms used to calculate correlation functions of elementary fields. From this perspective, the strategy we have outlined in the previous sections appears to achieve this. However, there is an important issue we must address: the highly formal nature of the expressions we employed. We expressed all mathematical formulas in 
$d$ dimensions because we had Dimensional Regularization (DR) in mind. However, if we set $d=4$, all integrals corresponding to Feynman diagrams become divergent. The only way to resolve this issue is by renormalizing the theory. This is well-established; the challenge now is determining how to make the theory defined by $S_B$ consistent. \par 
 
Since the propagator of $B$ is constant, internal lines of $B$ do not improve the convergence of integrals. Furthermore, as we already discussed, one $B$ internal line can, effectively, be removed from a diagram and replaced by a $\phi^{2n}$-type vertex. This last observation indicates what happens with the theory characterized by $S_B$ regarding the renormalizability. If $\phi^{2n}$ is a non-renormalizable interaction, then the same applies to this theory with $B$. Therefore, the dimension of the composite operator $O\left(x\right)$ determines how complicated the other theory will be. Eventhough the theory defined by $S$ might be renormalizable, and the dimension of $\rho$ is non-negative, which means that only a finite number of counterterms with $\rho$ are necessary to renormalize all correlation functions of $O$, if $n>\frac{d}{2}$ we have to deal with a non-renormalizable theory defined by $S_B$. Such situation was somehow expected, since we mapped the original problem to a more complex theory. Fortunately, once again, the parameter $\alpha$ and Eq. \eqref{eq:master_equation} keep the problem under control. Since, for a certain correlation function $\langle B(x_1)\ldots B(x_r)\rangle$ we are not interested in the whole expansion in $\alpha$, but only the $r^{th}$-order, the number of relevant divergences should be the same. \par

To ilustrate what we just said in the final of the last paragraph, let us analyze the superficial degree of divergence of the diagrams in the $\lambda \phi^N$ theory with the composite operator $\frac{\phi^n\left(x\right)}{n!}$, and later compare with the theory defined by $S_B$. A diagram with $L$ loops and $I_{\phi}$ internal lines of $\phi$ has the following superficial degree of divergence $\delta$: 
\begin{gather}
	\delta=dL-2I_{\phi}. \label{eq:superficial_degree}
\end{gather}
The number of loops can be determined by the Euler's formula:
\begin{gather}
	L=I_{\phi}-E_{\phi^n}-V_{\phi^N}+1,
\end{gather}
where $E_{\phi^n}$ is the number of composite operator insertions and $V_{\phi^N}$ is the number of $\phi^N$ vertices in the diagram. As we already discussed, composite operators behaves like vertices, thus, we include them as such in the equation for the number of loops. The number of internal lines is determined by the number of contractions left between the fields of vertices after we contract some of them with the external fields. Therefore, it follows that
\begin{gather}
	I_{\phi}=\frac{nE_{\phi^n}+NV_{\phi^N}-E_{\phi}}{2}.
\end{gather}
For the theory defined by $S_B$, Eq. \eqref{eq:superficial_degree} holds, since the internal lines of $B$ do not improve the convergence of the integrals. The number of loops is modified to
\begin{gather}
	L=I_{\phi}+I_{B}-V_{B\phi^n}-V_{\phi^N}+1,
\end{gather}
where $I_B$ is the number of internal lines of $B$ and $V_{B\phi^n}$ is the number of $B\phi^n$-vertices that exist in the diagram. Now, we have two expressions for the internal lines, namely,
\begin{gather}
	I_{\phi}=\frac{nV_{B\phi^n}+NV_{\phi^N}-E_{\phi}}{2}, \nonumber \\
	I_{B}=\frac{V_{B\phi^n}-E_{B}}{2},
\end{gather}
where $E_{B}$ is the numeber of external lines of $B$. Here, we are interested in the situation where the number of $B\phi$-vertices is equal to the number of extenal lines of $B$, \emph{i.e.}, $V_{B\phi^n}=E_{B}$. Notice that, in this case, we recover the power counting formulas of the $\lambda\phi^N$ theory with the composite operator $\frac{\phi^n}{n!}$, as we have to indentify $E_{B}=E_{\phi^n}$, according to Eq. \eqref{eq:master_equation}. Therefore, it seems that the original problem is not changed, \emph{i.e.}, we have do deal with the same number of UV-divergences.

\section*{Acknowledgements}
The author is grateful to Antonio Duarte Pereira Junior for helpful discussions. The author would like to thank the Brazilian agency FAPERJ for financial support. G. Peruzzo is a FAPERJ postdoctoral fellow in the P\'{O}S-DOUTORADO NOTA 10 program under the contracts E-26/205.924/2022 and E-26/205.925/2022.

\section*{Appendix A: Alternative derivation of the Master Equation \eqref{eq:master_equation} via a functional equation}}

In Section \ref{sec:Introducing-the-new}, we derived the master equation \eqref{eq:master_equation}, which establishes the relationship between the correlation functions of $B\left(x\right)$ and those of the composite operator $O\left(x\right)$. This result follows directly from the analysis of the Feynman rules and the Feynman diagrams. However, there exists an alternative way of demonstrating it, which employs a functional equation (a Ward identity), that we will also present in this appendix. \par 

Let us introduce in the theory defined by the action $S_{B}$, see \eqref{eq:Sb}, the composite operator $O\left(x\right)$ coupled with the external source $\rho$. Hence, we will work with the modified action
\begin{gather}
	S'_{B}=S_{B}+\int d^4 x\, \rho\left(x\right)O \left(x\right).
\end{gather}
Notice that 
\begin{gather}
S'_{B}|_{\rho=0}=S_{B}, \nonumber \\
 S'_{B}|_{\alpha=0}=S+\int d^4x \frac{B^2}{2}. \label{eq:limits}
\end{gather} 
In the latter case, since $B\left(x\right)$ is not an interacting field, $S'_{B}|_{\alpha=0}$ is equivalent to $S$.
\par 
The equation of motion of $B(x)$ can be written as
\begin{gather}
	\frac{\delta S'_{B}}{\delta B\left(x\right)}=B(x)-\alpha\frac{\delta S'_{B}}{\delta \rho \left(x\right)}, \label{eq:ward_b}
\end{gather}
which is a genuine Ward identity \cite{Piguet:1995er}. The Eq.\eqref{eq:ward_b} implies that the effective action $\Gamma'_B$ also satisfies the same identity, namely,
\begin{gather}
	\frac{\delta \Gamma'_{B}[\varphi,B,\rho,\alpha]}{\delta B\left(x\right)}=B(x)-\alpha\frac{\delta \Gamma'_{B}[\varphi,B,\rho,\alpha]}{\delta \rho \left(x\right)}. \label{eq:ward_b_gamma}
\end{gather}
The generating functional of the connected correlation functions and the effective action are related by a Legendre transformation:
\begin{gather}
	W'_B[J_\varphi,J_B,\rho,\alpha]=\Gamma'_B[\varphi,B,\rho,\alpha]+\int d^4x\,\left(J_\varphi \varphi+J_B B\right),
\end{gather}
where $J_\varphi$ and $J_B$ are the external sources of $\varphi$ and $B$, respectively. $W$ and the generating functional of the correlation functions, see Eq.\eqref{eq:Z_functional}, are related by the equation $Z=e^{-W}$. Therefore, the Ward identity \eqref{eq:ward_b_gamma} can be re-expressed as:
\begin{gather}
	-J_B\left(x\right)=\frac{\delta W'_B[J_\varphi,J_B,\rho,\alpha]}{\delta J_B\left(x\right)}-\alpha\frac{\delta W'_{B}[J_\varphi,J_B,\rho,\alpha]}{\delta \rho \left(x\right)}. \label{eq:ward_b_W}.
\end{gather}
By taking an appropriate number of derivatives with respect to $J_B$, $\rho$ and $\alpha$, we obtain \eqref{eq:master_equation}. Using the simple argument of continuity, and according to Eq. \eqref{eq:limits}, we must have 
\begin{gather}
	W'_{B}[J_\varphi,J_B,0,\alpha]=W_{B}[J_\varphi,J_B,\alpha], \nonumber \\
	W'_{B}[J_\varphi,0,\rho,0]=W[J_\varphi,\rho],
	\end{gather}
	where $W_B[J_\varphi,J_B,\alpha]$ and $W[J_\varphi,\rho]$  are the generating functionals associated with the actions $S_{B}$ and $S$, respectively. \par
Let us consider the example of the one-point function. By acting with $\partial/\partial \alpha$ on \eqref{eq:ward_b_W}, we obtain
\begin{gather}
	0=\frac{\partial}{\partial \alpha}\frac{\delta W'_B[J_\varphi,J_B,\rho,\alpha]}{\delta J_B\left(x\right)}-\frac{\delta W'_{B}[J_\varphi,J_B,\rho,\alpha]}{\delta \rho \left(x\right)}-\alpha\frac{\partial}{\partial \alpha}\frac{\delta W'_{B}[J_\varphi,J_B,\rho,\alpha]}{\delta \rho \left(x\right)}. \label{eq:ward_b_W2}.
\end{gather}
Now, by setting $\alpha$ and all other sources to zero, it follows that
\begin{gather}
	0=\left.\frac{\partial}{\partial \alpha}\langle B\left(x\right)\rangle^{S_B}_c\right|_{\alpha=0}-\langle O\left(x\right)\rangle^S_c,\label{eq:one_point}
\end{gather}
since it is also true that
\begin{gather}
	\left. \frac{\delta W'_{B}[J_\varphi,J_B,\rho,\alpha]}{\delta \rho \left(x\right)} \right| _{\alpha=J=\rho=0}=\left. \frac{\delta W'_{B}[0,0,\rho,0]}{\delta \rho \left(x\right)} \right| _{\rho=0}=\left. \frac{\delta W[J_\varphi,\rho]}{\delta \rho \left(x\right)} \right| _{J_\varphi=\rho=0}=\langle O\left(x\right)\rangle^S_c
\end{gather}
and
\begin{gather}
\left.	\frac{\partial}{\partial \alpha}\frac{\delta W'_B[J_\varphi,J_B,\rho,\alpha]}{\delta J_B\left(x\right)}\right|_{\alpha=J=\rho=0}=\left.	\frac{\partial}{\partial \alpha}\frac{\delta W'_B[0,J_B,0,\alpha]}{\delta J_B\left(x\right)}\right|_{\alpha=J_B=0}=\left.	\frac{\partial}{\partial \alpha}\frac{\delta W_B[J_\varphi,J_B,\alpha]}{\delta J_B\left(x\right)}\right|_{\alpha=J=0}=\left.\frac{\partial}{\partial \alpha}\langle B\left(x\right)\rangle^{S_B}_c\right|_{\alpha=0}, \nonumber\\
\end{gather}
 Notice that \eqref{eq:one_point} is indeed equivalent to \eqref{eq:master_equation} for $r=1$.


\end{document}